\documentstyle[11pt,newpasp,twoside,epsf]{article}
\def\etal{et~al.}
\def\ga{\stackrel{>}{\sim}}
\def\la{\stackrel{<}{\sim}}

\def\edcomment#1{\iffalse\marginpar{\raggedright\sl#1\/}\else\relax\fi}
\reversemarginpar

\begin{document}
\title{Statistical Properties of DLAs and sub-DLAs}
\author{C\'eline P\'eroux, Mike J. Irwin \& Richard G. McMahon}
\affil{Institute of Astronomy, Madingley Road, Cambridge CB3 0HA, UK}
\author{Lisa J. Storrie-Lombardi}
\affil{SIRTF Science Center, CalTech, Pasadena, USA}

\begin{abstract}
Quasar absorbers provide a powerful observational tool with which to
probe both galaxies and the intergalactic medium up to high
redshift. We present a study of the evolution of the column density
distribution, $f(N,z)$, and total neutral hydrogen mass in high-column
density quasar absorbers using data from a recent high-redshift survey
for damped Lyman-$\alpha$ (DLA) and Lyman limit system (LLS)
absorbers. Whilst in the redshift range 2 to 3.5, $\sim$90\% of the
neutral HI mass is in DLAs, we find that at z$>$3.5 this fraction
drops to only 55$\%$ and that the remaining 'missing' mass fraction of
the neutral gas lies in sub-DLAs with N(HI) $\rm 10^{19} - 2 \times
10^{20}~cm^{-2}$.
\end{abstract}

\section{Introduction}
Intervening absorption systems in the spectra of quasars provide a
unique way to study early epochs and galaxy progenitors. In
particular, they are not affected by the ``redshift desert'' from $
1.3 \stackrel{<}{_{\sim}} z \stackrel{<}{_{\sim}} 2.5 $ where spectral
emission features in normal galaxies do not fall in optical passbands,
yet where substantial galaxy formation is taking place. In addition,
the absorbers are selected strictly by gas cross-section, regardless
of luminosity, star formation rate, or morphology. Quasar absorbers
are divided according to their neutral hydrogen column densities: DLAs
have N(HI) $>2 \times 10^{20}$ cm$^{-2}$, Lyman-limit systems have
N(HI) $>1.6 \times 10^{17}$ cm$^{-2}$ and any system below this
threshold is known as the Lyman-$\alpha$ forest. The analysis
presented here is based on a sample of quasar absorbers found in the
spectra of 66 $z \ga 4$ quasars (Peroux et al. 2001b) combined with
data from the literature (Storrie-Lombardi \& Wolfe 2000).
 
\section{Quasar Absorbers Number Density and Column Density Distribution}

The absorption lines evolution is usually described with a power law
of the form: $n(z) dz = n_0 (1+z)^{\gamma} dz$, where $n(z)$ is the
observed number density of absorbers.  For convenience, the $dz$ is
dropped and the differential number density per unit redshift is
expressed as follow: $n(z) = n_0 (1+z)^{\gamma}$.  The observed number
density of absorbers is the product of the space density and physical
cross-section of the absorbers which are a function of the geometry of
the Universe. For no evolution of the properties of the individual
aborbers in a ${\Lambda}=0$ Universe, this yields $\gamma=1$ for
$\Omega_{M}=0$ and $\gamma=0.5$ for $\Omega_M =1$. We found $\gamma =
2.45^{+0.75}_{-0.65}$ indicating evolution at the $\sim 2 \sigma$
level independently of $\Omega_M$. Figure 1 shows the number density
of various classes of quasar absorbers.

\begin{figure}
\plotfiddle{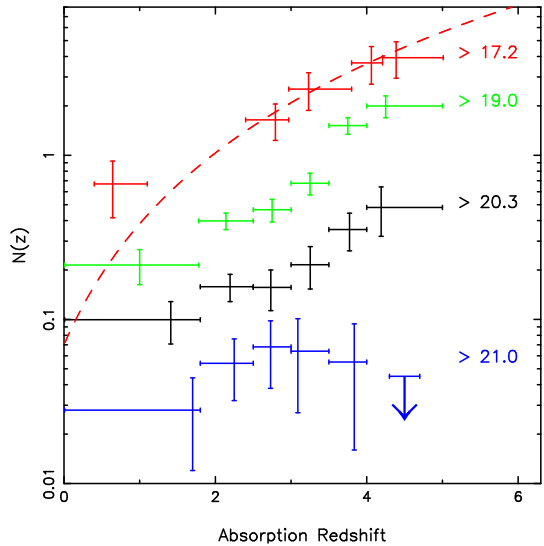}{7cm}{0}{40}{40}{-130}{-15}
\caption{Number density of quasar absorbers with (from top to bottom)
log $N(HI) > 17.2$, $>19.0$, $>20.3$ and $>21.0~cm^{-2}$. The column
density range $10^{19}< N(HI) < 10^{20.3}~cm^{-2}$ are {\it not}
direct observations but re-computed from the $\Gamma$-fit to the
column density distribution including the expected number of LLS. The
dashed line is a power-law fit to the number density of LLS with
$z_{abs} > 2.4$. No absorber with log $N(HI) > 21.0~cm^{-2}$ are
observed at $z>4$ and the arrow indicates the $50\%$ confidence upper
limit.}
\end{figure}

The column density distribution is determined as follow:

\begin{equation}
f(N, z) dN dX = \frac{n}{\Delta N \sum_{i=1}^{m} \Delta X_i} dN dX
\end{equation}

where $n$ is the number of quasar absorbers observed in a column
density bin $[N, N+dN]$ obtained from the observation of $m$ quasar
spectra with total absorption distance coverage $\sum_{i=1}^{m} \Delta
X_i$. The distance interval, $dX$, is used to correct to co-moving
coordinates and thus depends on the geometry of the Universe. In a
non-zero $\Lambda$-Universe:

\begin{equation}
X(z) = \int_{0}^{z} (1 + z)^2 \left[(1 + z)^2 (1 +
z\Omega_M) - z (2 + z) \Omega_{\Lambda}\right]^{-1/2}dz
\end{equation}

The column density distribution was usually fitted by a power law over
a large column density range $N(HI)$ $10^{12} - 10^{21}$
cm$^{-2}$. This suggested that all classes of absorbers arise from the
same cloud population (Tytler 1987). Assuming randomly distributed
spherical isothermal halos, $f(N,z)$ was well approximated with a
power law of slope $-5 /3$ (Rees 1988). Nevertheless, as the quality
of the data increased, deviations from a power law have been
observed. In particular Petitjean et al. (1993) observed a change in
slope: $\beta = -1.83$ for $N(HI) \leq 10^{16}$ and $\beta = -1.32$
above that threshold (Figure 2).

\begin{figure}
\plotfiddle{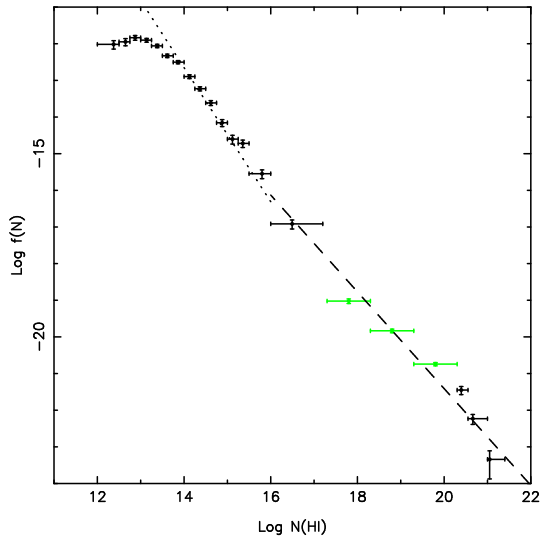}{7cm}{0}{40}{40}{-130}{-15}
\caption{Column density distribution, $f(N,z)$, at $z_{abs} >
3.5$. The low column density data are Keck-HIRES observations of the
Lyman-$\alpha$ forest (BR 1033$-$0327 and Q0000$-$26, Williger \etal\
1994 and Lu \etal\ 1996, respectively). The light grey bins (in the
range $17.2 < $ log $ N(HI) < 20.3$) are deduced from the fit to the
observed cumulative number of quasar absorbers. The turn-over at the
low column density end is incompleteness due to a combination of
spectral resolution and signal-to-noise. The dashed and dotted lines
are the two $<z> \simeq 2.8$ power law fits from Petitjean \etal\
(1993) corrected for the absorber number density evolution with
redshift and to $\Omega_{\Lambda}=0.7$, $\Omega_{M}=0.3$ cosmology.}
\end{figure}

However, LLS line profiles cannot be used to directly measure their
column densities in the range $10^{17.2}$ to $10^{19.0}$ cm$^{-2}$
because the curve of growth is degenerate in that interval. In our
analysis, we use the {\it expected number} of LLS to provide a further
constraint on the cumulative number of quasar absorbers and as clear
evidence that a simple power law does {\it not} fit the observations:

\begin{equation}
LLS_{expected}= \sum_{i=1}^{n}
\int_{z_{min}}^{z_{max}}N_o(1+z)^{\gamma}dz
\end{equation}

where $z_{min}$ and $z_{max}$ is the redshift path along which quasar
absorbers were searched for. We choose to fit the data with a
$\Gamma$-distribution (a power law with an exponential turn-over)
which was introduced by Pei $\&$ Fall (1995) and Storrie-Lombardi,
Irwin \& McMahon (1996b):

\begin{equation}
f(N,z) = (f_*/N_*)(N/N_*)^{-\beta} e^{-N/N_*} 
\end{equation}

where $N$ is the column density, $N_*$ a characteristic column density
and $f_*$ a normalising constant. Figure 3 shows the differential
column density distribution of quasar absorbers with the
$\Gamma$-distribution fit for various redshift ranges. The redshift
evolution indicates that there are less high column density systems at
high-redshift than at low redshift, confirming the earlier results
from Storrie-Lombardi, McMahon \& Irwin (1996a) and Storrie-Lombardi
\& Wolfe (2000). This suggests that we are observing the epoch of
formation of DLAs.

\begin{figure}
\plotfiddle{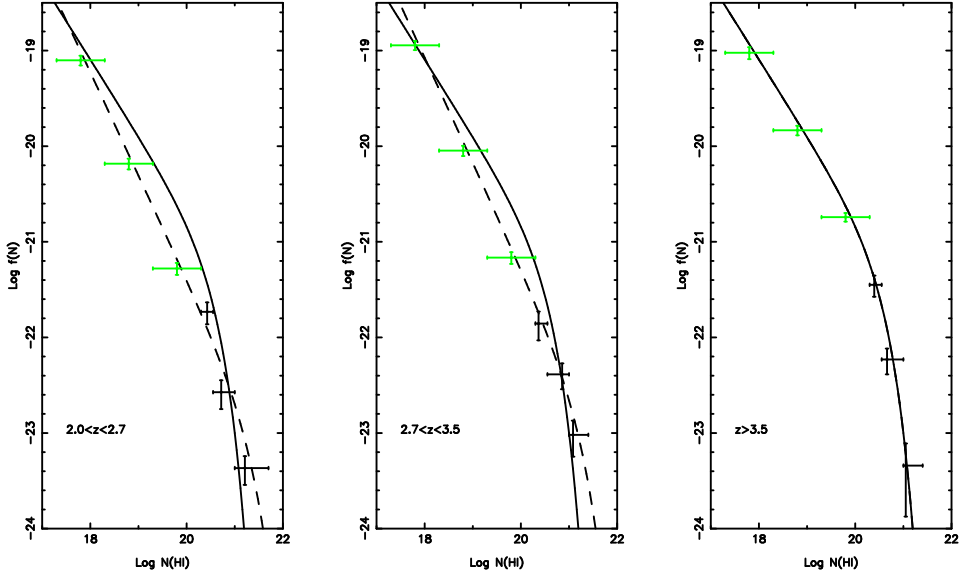}{8cm}{-90}{50}{50}{-200}{+250}
\caption{Column density distribution of quasar absorbers for various
redshift ranges. The redshift ranges are chosen to match the bins in
the $\Omega_{HI+HeII}$ plot (see Figure 4). The light grey bins (in
the range $17.2 < $ log $ N(HI) < 20.3$) are deduced from the fit to
the observed cumulative number of quasar absorbers. The solid line is
the $\Gamma$-distribution fit for $z > 3.5$ and the dashed lines are
the fits to the unbinned data in the relevant redshift range.}
\end{figure}

\section{Cosmological Evolution of Neutral Gas Mass}

The mass density of absorbers can be expressed in units of the current
critical mass density, $\rho_{crit}$, as:

\begin{equation}
\Omega_{HI+HeII}(z) = \frac{H_o \mu m_H}{c \rho_{crit}}
\int_{N_{min}}^{\infty} N f(N,z) dN
\end{equation}

where $\mu$, the mean molecular weight is $1.3$ and $m_H$ is the
hydrogen mass. The total HI may be estimated as:

\begin{equation}
\int_{N_{min}}^{\infty} N f(N,z) dN = \frac{\sum N_i(HI)}{\Delta X}
\end{equation}

If a power law is used to fit $f(N,z)$, up to $90\%$ of the neutral
gas is in DLAs (Lanzetta, Wolfe \& Turnshek 1995), although an
artificial cut-off needs to be introduced at the high column density
end because of the divergence of the integral. If instead a
$\Gamma$-distribution is fitted to $f(N,z)$ this removes the need to
artificially truncate the high end column distribution and can be used
to probe in more detail the neutral gas fraction as a function of
column density and how this changes with redshift. The resulting
$\Omega_{HI+HeII}$ is shown in Figure 4.

\begin{figure}
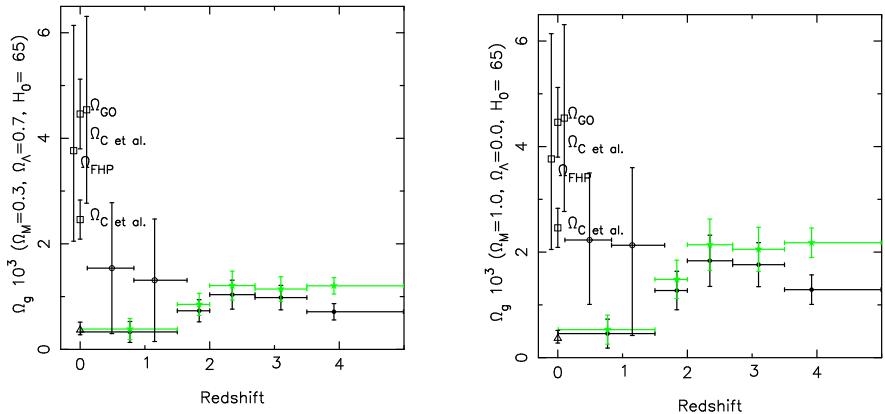

\plotfiddle{peroux_fig4.eps}{2.5cm}{0}{32}{32}{-180}{-100}
\plotfiddle{peroux_fig5.eps}{2.5cm}{0}{32}{32}{0}{-19}
\caption{The black circles show the neutral gas in Damped
Lyman-$\alpha$ galaxies in a $\Omega_{\Lambda}=0.7$, $\Omega_{M}=0.3$
and $h=0.65$ Universe (left panel). Vertical error bars correspond to
1-$\sigma$ uncertainties and the horizontal error bars indicate bin
sizes. The light grey stars are the total HI+HeII including a
correction for the neutral gas contained in sub-DLAs. The circles at
low redshift are the measurements from Rao $\&$ Turnshek (2000), who
used a method involving the observations of quasar spectra with known
MgII systems. The triangle at $z=0$ is the local HI mass measured by
Natarajan $\&$ Pettini (1997). The squares, $\Omega_{FHP}$,
$\Omega_{GO}$ and $\Omega_{C et al.}$ (Fukugita, Hogan $\&$ Peebles
1998, Gnedin $\&$ Ostriker 1992 and Cole \etal\ 2000 respectively) are
$\Omega_{baryons}$ in local galaxies. The right panel is for a
$\Omega_{M} = 1$ cosmology. This plot shows that the geometry of the
Universe affects the absolute value of $\Omega_{HI+HeII}$ with respect
to the local $\Omega_{baryons}$.}
\end{figure}

\section{Results and Discussion}

We find that at z$>$3.5 the fraction of mass in DLAs is only 55$\%$
and that the remaining fraction of the neutral gas mass lies in
systems below this limit, in the so-called ``sub-DLAs'' with column
density $10^{19} <$ N(HI) $< 2 \times 10^{20}$ cm$^{-2}$ (Peroux et
al. 2001a). Our observations in the redshift range 2 to 5 are
consistent with no evolution in the {\it total} amount of neutral
gas. Under simple assumptions of closed box evolution, his could be
interpreted as indicating there is little gas consumption due to star
formation in DLA systems in this redshift range. Similarly, at z $>$
2, Prochaska, Gawiser \& Wolfe (2001) conclude that there is no
evolution in the metallicity of DLA systems from column
density-weighted Fe abundance measurements in DLAs (see also Savaglio
2000). At low-redshift, recent measurements of $\Omega_{HI+HeII}$ by
Rao \& Turnshek (2000) at $z \la 1.65$ and Churchill (2001) at $z \sim
0.05$ are difficult to reconcile with 21 cm emission observations at
$z=0$. Nevertheless, the cosmological evolution of the {\it total}
neutral gas mass proves to be a powerful way of tracing galaxy
formation with redshift: it probes the epoch of assembly of high
column density systems from lower column density units.

\acknowledgments CP would like to thank the organising committee for
putting together a very enjoyable meeting.

\end{document}